\documentclass[12pt]{iopart}

\usepackage{iopams}

\newcommand{\kommentar}[1]{}

\newcommand{\ud}{{\rm d}}
\newcommand{\ui}{{\rm i}}
\newcommand{\ue}{{\rm e}}

\newcommand{\R}{{\mathbb R}}

\newcommand{\Z}{{\mathbb Z}}

\newcommand{\vecp}{\vec{p}}
\newcommand{\vecx}{\vec{x}}

\begin{document}


\noindent
ULM-TP/00-4 \\
September 2000\\

\jl{1}

\title[Two-point correlations of the GSE from periodic orbits]{Two-point 
correlations of the Gaussian symplectic ensemble from periodic orbits}

\author{Stefan Keppeler%
\footnote{E-mail address: {\tt kep@physik.uni-ulm.de}}%
}

\address{Abteilung Theoretische Physik, Universit\"at Ulm, 
Albert-Einstein-Allee 11, D-89069 Ulm, Germany}

\begin{abstract}
We determine the asymptotics of the two-point correlation function for
quantum systems with half-integer spin which show chaotic behaviour in the
classical limit using a method introduced by Bogomolny and Keating 
[Phys.~Rev.~Lett. {\bf 77}~(1996)~1472--1475]. 
For time-reversal invariant systems we obtain the leading
terms of the two-point correlation function of the Gaussian symplectic 
ensemble. Special attention has to be paid to the r\^ole of Kramers' 
degeneracy.
\end{abstract}



\maketitle

Understanding correlations of energy levels of quantum mechanical systems 
whose classical limit exhibits chaotic motion is one of the major topics in 
quantum chaos. The bridge between quantum mechanics and classical mechanics is
provided by the Gutzwiller trace formula \cite{Gut71} which relates the 
quantum mechanical density of states $d(E) = \sum_n \delta(E-E_n)$ to a sum 
over periodic orbits of the corresponding classical system,
\begin{equation}
\label{trace_d}
  d(E) \sim \bar{d}(E) 
  + \frac{1}{2\pi\hbar} \sum_{\gamma} \sum_{k\in\Z\setminus\{0\}} 
    \mathcal{A}_{\gamma k} \, T_{\gamma} \, 
    \ue^{\frac{\ui}{\hbar}k S_{\gamma}(E)} 
  \, , \quad  \hbar \to 0 \, ,
\end{equation}
where $\bar{d}(E)$ denotes the mean spectral density (which, by Weyl's law,  
is of order 
$\hbar^{-f}$ for systems with $f$ degrees of freedom), and the sum extends 
over all primitive periodic orbits $\gamma$ and their repetitions, formally 
including negative ones. $S_{\gamma}(E) = \oint_{\gamma} \vecp \, \ud \vecx$ 
denotes the classical action, $T_{\gamma}$ is the (primitive) period, 
$T_{\gamma} = \ud S_{\gamma}(E)/\ud E$, 
and the amplitude $\mathcal{A}_{\gamma k}$ involves topological and stability 
properties. The conjecture of Bohigas, Giannoni and Schmit (BGS) 
\cite{BohGiaSch84} states that for classically chaotic systems, generically,
the statistics 
of energy levels can be modeled by the average behaviour of ensembles of 
random matrices. In the case of systems without spin the relevant ensembles are
the Gaussian orthogonal and the Gaussian unitary ensemble (GOE/GUE) depending
on whether the system does or does not possess an antiunitary symmetry like 
time-reversal, see, e.g. \cite{Haa00}. In the
case of time-reversal invariant systems with half-integer spin one also has to
deal with the Gaussian symplectic ensemble (GSE). 

The main result in understanding eigenvalue correlations in terms of the 
underlying classical dynamics is due to Berry \cite{Ber85}. He used the 
so-called diagonal approximation and the Hannay-Ozorio de Almeida sum rule
\cite{HanOzo84}, see also \cite{ParPol90}, to determine the asymptotics of the
spectral form factor, which is the Fourier transform of the two-point 
correlation function $R_2(s)$, see eq. (\ref{R2def}) below. This treatment has 
recently been generalized to the case with half-integer spin \cite{BolKep99b}
using an analogue of the Gutzwiller trace formula which includes an
additional factor due to spin precession \cite{BolKep98,BolKep99a}. 

In the case of the GOE and the GUE Bogomolny and Keating \cite{BogKea96}, see
also \cite{Bog98,Kea99}, developed a method for the semiclassical evaluation 
of $R_2(s)$ which yields an additional term as compared to the diagonal 
approximation of the form factor. More precisely, their method yields the 
leading non-oscillatory and the leading oscillatory term of $R_2(s)$ as 
$s\to\infty$, whereas the diagonal approximation of the form factor corresponds
to the leading non-oscillatory term.
Recently Haake \cite{Haa00} proposed a method to adapt this result to the case
of the GSE. But, surprisingly, although he obtained two terms of the large $s$
asymptotics of $R_2(s)$, the method failed to reproduce the leading term. 
The aim of this Letter is to present a slightly different approach to 
systems with half-integer spin which correctly yields both the leading 
non-oscillatory and the leading oscillatory term.
Note that \cite{BogKea96} also includes a remark on GSE asymptotics which,
however, is not based on semiclassics with spin but on a theorem in random 
matrix theory and, therefore, is not to be confused with the problem 
addressed here.

The general method of \cite{BogKea96} consists of three main steps. Starting 
from the 
observation that trace formulae lead to accurate semiclassical quantization 
conditions Bogomolny and Keating propose to base the semiclassical analysis of 
spectral correlations on such an approximate spectrum. In the course of the 
calculations they, secondly, employ the diagonal approximation as introduced 
in \cite{HanOzo84,Ber85}. Finally, they make use of the assumption that the 
oscillating part of the integrated spectral density (i.e. the contribution of 
periodic orbits) behaves like a Gaussian random variable. 
Here we will only briefly sketch the necessary changes to the method of 
Bogomolny and Keating in order to take care of the situation with half-integer 
spin. For the general formalism we refer to \cite{BogKea96,Bog98,Kea99,Haa00}. 
We will also rely heavily on results of \cite{BolKep99b}.

In order to obtain a simple but efficient semiclassical quantization condition,
we first integrate (\ref{trace_d}) over the 
energy $E$ which yields a trace formula for the spectral staircase function 
$N(E)$. Taking into account only orbits up to a time $T$, which below will be 
chosen of the order of Heisenberg time $T_H = 2\pi\hbar\bar{d}(E)$, one 
obtains a truncated spectral staircase function $N_T(E)$, and the 
semiclassical eigenvalues $E_n(T)$ can be determined from the condition 
\cite{AurSte92a, AurMatSieSte92}
\begin{equation}
\label{cosquant}
  N_T(E_n(T)) \stackrel{!}{=} n + \frac{1}{2} \, .
\end{equation}
The trace formula (\ref{trace_d}) can easily be integrated if there is a 
one-to-one correspondence between orbits at different energies 
(i.e. no bifurcations occur when varying $E$) and from successive 
integration by parts we see that in leading 
order in $\hbar$ it is sufficient to integrate the exponential, i.e 
\begin{equation}
\label{trace_N}
  N_T(E) \sim \bar{N}(E) 
  + \sum_{\gamma} \sum_{k\in\Z\setminus\{0\} \atop |k|T_\gamma \leq T }  
    \frac{1}{2\pi\ui k} \, \mathcal{A}_{\gamma k} \, 
    \ue^{\frac{\ui}{\hbar}k S_{\gamma}(E)} 
  \, , \quad  \hbar \to 0 \, ,
\end{equation}
where the periodic orbit sum will later be denoted by $N_T^{\rm fluc}(E)$.
At this point it is important to take care of Kramers' degeneracy. If the 
quantum system, with Hamiltonian $\hat{H}$, has half-integer spin and is 
invariant under time-reversal, i.e. 
$[\hat{H},\hat{T}]=0$ with $\hat{T}=\ui\sigma_y\hat{K}$, where $\hat{K}$ is
the operator of complex conjugation, then each energy eigenvalue has at least 
multiplicity two. One could now attempt to first calculate the correlations for
the degenerate spectrum and relate the result to the correlations of the 
non-degenerate spectrum, cf. \cite{Haa00}. This strategy is successful for the 
form factor \cite{BolKep99b}. However, since the truncated spectral straicase 
function $N_T(E)$ fails to reproduce sharp steps of size two, the quantization 
condition (\ref{cosquant}) will not yield degenerate eigenvalues but two 
distinct eigenvalues which both have an additional error. Therefore,
we instead take Kramers' degeneracy into account at this point by imposing
the modified quantization condition
\begin{equation}
  N_T(E_n(T)) \stackrel{!}{=} 2n + 1
\end{equation}
which produces a semiclassical spectrum $\{ E_n(T) \}$ with Kramers' degeneracy
already removed. Note that this semiclassical spectrum has mean density 
$\bar{d}(E)/2$, and the corresponding Heisenberg time is 
$T_H=\pi\hbar\bar{d}(E)$. Using the Poisson summation formula, the density of 
states $\tilde{d}(E)$ of the semiclassical spectrum can be written as
\begin{equation}
\label{dtilde}
  \tilde{d}(E) := \sum_n \delta(E-E_n(T)) 
  = \frac{1}{2} d_T(E) \sum_{\nu\in\Z} (-1)^\nu \, \ue^{\ui\pi\nu N_T(E)} \, ,
\end{equation}
where $d_T(E) = \ud N_T(E)/\ud T$, see (\ref{trace_N}).
Before we can compare spectral correlations with results from random matrix 
theory (RMT) the spectrum has to be unfolded, i.e. the eigenenergies are 
rescaled such that their mean separation is one. To this end consider the
spectral interval $I=I(E,\hbar):=[E-\hbar\omega,E+\hbar\omega]$ which contains
$N_I$ levels. In the semiclassical limit this number can be estimated by 
$N_I \sim 2\hbar\omega \bar{d}/2$, where $\bar{d} = \bar{d}(E)$, i.e. as 
$\hbar \to 0$ the length of the 
interval shrinks to zero but the number of eigenvalues contained in 
$I(E,\hbar)$ goes to infinity, cf. \cite{BolKep99b}. Defining the 
unfolded eigenvalues by $x_n(T):=E_n(T) \bar{d}/2$, the density of states 
$D_T(x)$, $x=E\bar{d}/2$, of the unfolded spectrum $\{x_n(T)\}$ reads 
\begin{equation}
\label{Ddef}
  D_T(x):= \sum_n \delta(x-x_n(T)) = \frac{2}{\bar{d}} \, \tilde{d}_T(E) \, .
\end{equation}
The semiclassical two-point correlation function is defined by
\begin{equation}
\label{R2def}
  R_2(s,I)
   := \frac{1}{\hbar\omega\bar{d}} 
    \int_{x-\hbar\omega\bar{d}/2}^{x+\hbar\omega\bar{d}/2} 
    D_T \left( x^\prime + \frac{s}{2} \right) \, 
    D_T \left( x^\prime - \frac{s}{2} \right) \, \ud x^\prime - 1 \, .
\end{equation}
From the BGS-conjecture we expect that in the semiclassical limit $R_2(s,I)$ 
converges weakly to the random matrix result ($R_2^{\rm GSE}(s)$ in the case
considered here), i.e.
\begin{equation}
  \lim_{\hbar\to 0} \int_{\R} R_2(s,I) \, \phi(s) \, \ud s
  = \int_{\R} R_2^{\rm GSE}(s) \, \phi(s) \, \ud s
\end{equation}
for any smooth test function $\phi \in \mathcal{S}(\R)$. We only aim at 
providing a periodic orbit theory for this relation in the combined limit
\begin{equation}
\label{limit}
  s \to \infty \, , \quad \bar{d} \to \infty \quad {\rm and} \quad
  s/\bar{d} \to 0
\end{equation}
which will allow expansions in $s/\bar{d}$. Here $\bar{d} \to \infty$ 
is a consequence of the semiclassical limit and Weyl's law. 
The asymptotics of the GSE-result reads (see, e.g. \cite{Meh91})
\begin{equation}
\label{RGSE}
  R_2^{\rm GSE}(s) \sim 
  \frac{\pi}{2} \frac{\cos(2\pi s)}{2\pi s} 
  - \frac{1+\frac{\pi}{2}\sin(2\pi s)}{(2\pi s)^2} \, , \quad s \to \infty \, .
\end{equation}
Substituting (\ref{Ddef}) and 
(\ref{dtilde}) into (\ref{R2def}) results in
\begin{equation}
\fl
  R_2(s,I) = 
  \frac{1}{\bar{d}^2} \Bigg\langle \!
  d_T \! \left( \textstyle E^\prime \! + \! \frac{s}{\bar{d}} \right) \, 
  d_T \! \left( \textstyle E^\prime \! - \! \frac{s}{\bar{d}} \right)
  \! \sum_{\nu,\nu^\prime \in \Z} \! (-1)^{\nu-\nu^\prime} 
  \ue^{\ui\pi \left( \nu N_T(E^\prime + \frac{s}{\bar{d}}) 
                     - \nu^\prime N_T(E^\prime-\frac{s}{\bar{d}}) \right)} 
  \!\Bigg\rangle_{\!\!E^\prime} -1 \, ,
\end{equation}
where the brackets denote an average over $I(E,\hbar)$, i.e. 
$\langle ... \rangle_{E^\prime} = \frac{1}{2\hbar\omega} 
 \int_{E-\hbar\omega}^{E+\hbar\omega} ... \, \ud E^\prime$. By a stationary 
phase argument one easily sees that the terms with $\nu \neq \nu^\prime$ are 
of relative order $\Or(1/\bar{d}\,)$ in the desired limit 
(\ref{limit}), i.e. we have 
\begin{equation}
  R_2(s,I) \sim \sum_{\nu\in\Z} r_\nu(s,I)
\end{equation}
with 
\begin{equation} 
\fl
  r_\nu(s,I) := \frac{1}{\bar{d}^2} \Bigg\langle
  d_T \left( \textstyle E^\prime + \frac{s}{\bar{d}} \right) \, 
  d_T \left( \textstyle E^\prime - \frac{s}{\bar{d}} \right)
  \ue^{\ui\pi \nu \left( N_T(E^\prime + \frac{s}{\bar{d}}) 
                       - N_T(E^\prime-\frac{s}{\bar{d}}) \right)} \, 
  \Bigg\rangle_{\!\!E^\prime} - \delta_{\nu 0} \, .
\end{equation}
The evaluation of $r_0(s,I)$ is straight forward and will not be shown here. 
The result corresponds to the usual diagonal approximation of the form factor
(cf. \cite{BogKea96,Haa00}) and therefore in the 
present situation reads \cite{BolKep99b}
\begin{equation}
\label{Rdiag}
  r_0(s,I) \approx - \frac{1}{(2\pi s)^2} \, ,
\end{equation}
which is the leading non-oscillating contribution of $R_2^{\rm GSE}(s)$ as 
$s\to \infty$ (\ref{RGSE}). Here `$\approx$' indicates that (\ref{Rdiag}) is 
not just an asymptotic relation but we have also used the diagonal 
approximation which is assumend to be valid in the combined limit 
(\ref{limit}).
Further conditions needed to arrive at (\ref{Rdiag}) are hyperbolicity of 
the translational dynamics and the mixing property of the skew product of 
translational and spin dynamics, see \cite{BolKep99b} for details. We remark 
that the last condition can be weakened to ergodicity; this result will be 
presented elsewhere \cite{Kep_prepare}.
Introducing an auxiliary variable $s^\prime$, the contributions $r_\nu(s,I)$, 
$\nu \neq 0$, can be written as derivatives, 
\begin{equation}
\label{rnuDef}
  r_\nu(s,I) \sim \frac{1}{(\ui\pi \nu)^2} \left.
  \frac{\partial^2}{\partial s\partial s^\prime} \, 
  \ue^{\ui\pi \nu (s+s^\prime)} \, 
  \Phi_\nu(s,s^\prime) \right|_{s^\prime=s} \, ,
\end{equation}
where we have expanded the smooth part $\bar{N}(E)$ of $N_T(E)$ in powers of 
$s/\bar{d}$. The functions $\Phi_\nu(s,s^\prime)$ are then defined by
\begin{equation}
\label{PhiDef}
  \Phi_\nu(s,s^\prime):= \left\langle
  \ue^{\ui\pi \nu \left( N_T^{\rm fluc}(E^\prime + \frac{s}{\bar{d}}) 
       - N_T^{\rm fluc}(E^\prime - \frac{s^\prime}{\bar{d}}) \right)}
  \right\rangle_{\!\!E^\prime} \, .
\end{equation}
The next step lies at the heart of the method of \cite{BogKea96}. Assuming 
that the exponent of (\ref{PhiDef}) behaves like a Gaussian random variable 
$G(E^\prime)$ with zero mean we can 
use the identity $\langle \exp(\ui G(E^\prime)) \rangle_{E^\prime} 
= \exp(\langle -G^2(E^\prime)/2 \rangle_{E^\prime})$ and subsequently evaluate 
the exponent in diagonal approximation. This assumption is favoured by a well 
established conjecture on global eigenvalue correlations \cite{AurBolSte94}.
Employing an expansion in $s/\bar{d}$ the difference in the exponent of 
(\ref{PhiDef}) reads 
\begin{equation}
\fl
  {\textstyle N_T^{\rm fluc}(E^\prime\!+\!\frac{s}{\bar{d}}) 
  - N_T^{\rm fluc}(E^\prime\!-\!\frac{s^\prime}{\bar{d}})
  } \sim
  \sum_{\gamma} \sum_{k\in\Z\setminus\{0\} \atop |k|T_\gamma \leq T} 
  \frac{\mathcal{A}_{\gamma k}}{2\pi\ui k} 
  \, \ue^{\frac{\ui}{\hbar}kS_{\gamma}(E^\prime)} 
  \left( \ue^{\frac{\ui}{\hbar}kT_{\gamma}(E^\prime)\frac{s}{\bar{d}}} 
    - \ue^{-\frac{\ui}{\hbar}kT_{\gamma}(E^\prime)\frac{s^\prime}{\bar{d}}} 
  \right) \, .
\end{equation}
For the square of this expression we again make use of the diagonal 
approximation, which was already needed to evaluate $r_0(s,I)$, i.e. we only 
keep contributions of orbits with like actions,
\begin{eqnarray}
\fl\nonumber
  \left\langle \left( {\textstyle 
  N_T^{\rm fluc}(E^\prime+\frac{s}{\bar{d}}) 
  - N_T^{\rm fluc}(E^\prime-\frac{s^\prime}{\bar{d}}) 
  } \right)^2 \right\rangle_{\!\!\!E^\prime}
  \approx \\ 
  \left\langle \sum_{\gamma} 
  \sum_{k\in\Z\setminus\{0\} \atop |k|T_\gamma \leq T} 
  \frac{g}{(2\pi k)^2} \, |\mathcal{A}_{\gamma k}|^2 \, 2 
  \left( 1 - \cos \left( \frac{k}{\hbar}T_{\gamma}(E^\prime) 
                         \frac{s+s^\prime}{\bar{d}} \right) \right)
  \right\rangle_{\!\!\!E^\prime} \, ,
\end{eqnarray}
where the generic multiplicity of periodic orbits, see, e.g., \cite{BolKep99b},
will be set to $g=2$, since we are dealing 
with time-reversal invariant systems. The remaining sum over periodic orbits 
can be evaluated with the sum rule of \cite{BolKep99b}, which, essentially, is
the Hannay-Ozorio de Almeida sum rule \cite{HanOzo84}, additionally 
taking into account the spin contribution. This yields
\begin{eqnarray}
\nonumber
\label{sumrule}
  \left\langle \left( {\textstyle 
  N_T^{\rm fluc}(E^\prime+\frac{s}{\bar{d}}) 
  - N_T^{\rm fluc}(E^\prime-\frac{s^\prime}{\bar{d}})
  } \right)^2 \right\rangle_{\!\!\!E^\prime} 
  &\approx \frac{g}{\pi^2} \int_0^T 
  \frac{1 - \cos\left( \frac{s+s^\prime}{\hbar\bar{d}} T^\prime \right)}
  {T^\prime} \, \ud T^\prime \\
  &\sim \frac{g}{\pi^2} \log \left( \frac{s+s^\prime}{\hbar\bar{d}} T \right)
  \, .
\end{eqnarray}
At this point a short remark is in order. The final result for $r_\nu(s,I)$, 
$\nu \neq 0$, will still include the cut-off time $T$. It has been argued 
\cite{BogKea96, Haa00} that $T$ has to be chosen of the order of Heisenberg 
time, i.e. $T=C \pi\hbar \bar{d}$, where the constant $C$ has to be 
determined by comparing to the asymptotics of the RMT-result. However, the 
result for $C$ will depend sensitively on the second term in the asymptotic 
expansion of the cosine-integral in (\ref{sumrule}). Terms of the same order 
could also arise from non-leading contributions to the sum rule, which, 
unfortunately, are unknown. We are thus unable to give the correct sub-leading 
term in the asymptotic expansion (\ref{sumrule}). 
Therefore we conclude that from the above 
considerations we can only obtain the $s$-dependence of $r_\nu(s,I)$, 
$\nu \neq 0$, but must refrain from any discussion about the cut-off time $T$.
Putting together (\ref{rnuDef}), (\ref{PhiDef}), (\ref{sumrule}) and the 
Gaussian ansatz we first observe that contributions with $|\nu| > 1$ decay 
rapidly. The leading correction to (\ref{Rdiag}) is hence given by 
$r_1(s,I)+r_{-1}(s,I)$, which yields a term proportional to $\cos(2\pi s)/s$.
This result is consistent with the leading oscillatory term of 
$R_2^{\rm GSE}(s)$, cf. (\ref{RGSE}).

Summarizing, the method of Bogomolny and Keating \cite{BogKea96} has been 
applied to time-reversal invariant systems with half-integer spin. As in the
previously studied cases without spin it correctly reproduces the leading 
non-oscillatory term and the $s$-dependence of the leading oscillatory term of 
the two-point correlation function as $s\to\infty$. Although giving further 
semiclassical evidence towards the BGS-conjecture open questions, as, e.g.,  
a consistent determination of the correct cut-off time $T$, remain.

\ack 

I would like to thank Prof.~F.~Haake for helpful suggestions and providing me 
with parts of his book manuscript \cite{Haa00} prior to publication and
Prof.~J.~P.~Keating for carefully discussing reference \cite{BogKea96} with
me and for many helpful remarks. I am also grateful to Dr.~J.~Bolte, 
R.~Schubert and Prof.~F.~Steiner for useful discussions. This work was partly 
supported by Deutscher Akademischer Austauschdienst (DAAD) under grant no. 
D/99/02553 and by Deutsche Forschungsgemeinschaft (DFG) under contract no. 
STE~241/10-1.

\section*{References}

\bibliographystyle{my_unsrt}
\bibliography{literatur}

\begin{thebibliography}{10}

\bibitem{Gut71}
M.~C. {G}utzwiller: {\em {P}eriodic {O}rbits and {C}lassical {Q}uantization
  {C}onditions\/}, J. Math. Phys. {\bf 12} (1971) ~343--358.

\bibitem{BohGiaSch84}
O.~{B}ohigas, M.-J. {G}iannoni and C.~{S}chmit: {\em {C}haracterization of
  chaotic quantum spectra and universality of level fluctuation laws\/}, Phys.
  Rev. Lett. {\bf 52} (1984) ~1--4.

\bibitem{Haa00}
F.~{H}aake: {\em {Q}uantum {S}ignatures of {C}haos\/}, {S}pringer-{V}erlag,
  {B}erlin {H}eidelberg,  (2000).

\bibitem{Ber85}
M.~V. {B}erry: {\em {S}emiclassical theory of spectral rigidity\/}, Proc. R.
  Soc. London Ser. A {\bf 400} (1985) ~229--251.

\bibitem{HanOzo84}
J.~H. Hannay and A.~M. {Ozorio de Almeida}: {\em Periodic orbits and a
  correlation function for the semiclassical density of states\/}, J. Phys. A
  {\bf 17} (1984) ~3429--3440.

\bibitem{ParPol90}
W.~{P}arry and M.~{P}ollicott: {\em Zeta functions and the periodic orbit
  structure of hyperbolic dynamics\/}, Ast\'erisque {\bf 187-8} (1990) ~1--268.

\bibitem{BolKep99b}
J.~{B}olte and S.~{K}eppeler: {\em {S}emiclassical form factor for chaotic
  systems with spin 1/2\/}, J. Phys. A {\bf 32} (1999) ~8863--8880.

\bibitem{BolKep98}
J.~{B}olte and S.~{K}eppeler: {\em {S}emiclassical Time Evolution and Trace
  Formula for Relativistic Spin-1/2 Particles\/}, Phys. Rev. Lett. {\bf 81}
  (1998) ~1987--1991.

\bibitem{BolKep99a}
J.~{B}olte and S.~{K}eppeler: {\em A semiclassical approach to the {D}irac
  equation\/}, Ann. Phys. (NY) {\bf 274} (1999) ~125--162.

\bibitem{BogKea96}
E.~B. Bogomolny and J.~P. Keating: {\em {G}utzwiller's {T}race {F}ormula and
  {S}pectral {S}tatistics: {B}eyond the {D}iagonal {A}pproximation\/}, Phys.
  Rev. Lett. {\bf 77} (1996) ~1472--1475.

\bibitem{Bog98}
E.~Bogomolny: {\em Spectral statistics\/}, in: {\em Proceedings of the
  International Congress of Mathematicians, (Berlin, 1998)\/},  99--108,
  (1998), Doc. Math. Extra Volume III.

\bibitem{Kea99}
J.~P. Keating: {\em Periodic Orbits, Spectral Statistics, and the Riemann
  Zeros\/}, in: {\em {S}uperymmetry and {T}race {F}ormulae: {C}haos and
  {D}isorder\/} (Eds. I.~V. Lerner, J.~P. Keating and D.~E. Khmelnitskii), vol.
  370 of {\em NATO ASI Series B: Physics\/},  1--15, New Yorck,  (1999), Kluwer
  Academic/Plenum Publishers.

\bibitem{AurSte92a}
R.~{A}urich and F.~{S}teiner: {\em {S}taircase functions, spectral rigidity,
  and a rule for quantizing chaos\/}, Phys. Rev. A {\bf 45} (1992) ~583--592.

\bibitem{AurMatSieSte92}
R.~{A}urich, C.~{M}atthies, M.~{S}ieber and F.~{S}teiner: {\em {N}ovel {R}ule
  for {Q}uantizing {C}haos\/}, Phys. Rev. Lett. {\bf 68} (1992) ~1629--1632.

\bibitem{Meh91}
M.~L. {M}ehta: {\em {R}andom {M}atrices\/}, Academic Press, San Diego, 2nd
  edn.,  (1991).

\bibitem{Kep_prepare}
S.~Keppeler: in preparation.

\bibitem{AurBolSte94}
R.~{A}urich, J.~{B}olte and F.~{S}teiner: {\em {U}niversal {S}ignatures of
  {Q}uantum {C}haos\/}, Phys. Rev. Lett. {\bf 73} (1994) ~1356--1359.

\end{thebibliography}

\end{document}